\documentclass[manuscript]{aastex}
\usepackage{graphicx}

\newcommand{\cu}{\object{CU\,Vir}}

\slugcomment{}

\shorttitle{"Auroral" Radio Emission from Cu Vir}
\shortauthors{Trigilio et al.}

\begin{document}

\title{Auroral Radio Emission from Stars:\\ the case of CU Virginis}


\author{Corrado Trigilio\altaffilmark{1}, Paolo Leto\altaffilmark{1}, Grazia Umana\altaffilmark{1}, Carla Simona Buemi\altaffilmark{1}, Francesco Leone\altaffilmark{2}}
\affil{$^{1}$ INAF-Osservatorio Astrofisico di Catania, Via S. Sofia 78, 95123
Catania, Italy\\
$^{2}$ Universit\`a di Catania, Dipartimento di Fisica e Astronomia, Via
Santa Sofia 78, 95123 Catania, Italy}
\email{ctrigilio@oact.inaf.it}


%\altaffiltext{1}{Visiting Astronomer, Cerro Tololo Inter-American Observatory.
%CTIO is operated by AURA, Inc.\ under contract to the National Science
%\altaffiltext{2}{Society of Fellows, Harvard University.}
%\altaffiltext{3}{present address: Center for Astrophysics,
%    60 Garden Street, Cambridge, MA 02138}
%\altaffiltext{4}{Visiting Programmer, Space Telescope Science Institute}
%\altaffiltext{5}{Patron, Alonso's Bar and Grill}


\begin{abstract}
CU Virginis is a rapidly rotating Magnetic Chemically Peculiar star with at present unique characteristics as radio emitter. The most intriguing one is the presence of intense, 100\% circularly polarized radiation ascribed to Cyclotron Maser. Each time the star rotates, this highly beamed emission points two times toward the Earth, like a pulsar. We observed CU Vir in April 2010 with the EVLA in two bands centered at 1450 and 1850 MHz. We covered nearly the whole rotational period, confirming the presence of the two pulses at a flux density up to 20 mJy. Dynamical spectra, obtained with unprecedented spectral and temporal sensitivity, allow us to clearly see the different time delays as a function of the frequency. We interpret this behaviour as a propagation effect of the radiation inside the stellar magnetosphere. The emerging scenario suggests interesting similarities with the auroral radio emission from planets, in particular with the Auroral Kilometric Radiation (AKR) from Earth, which originates at few terrestrial radii above the magnetic poles and was  only recently discovered to be highly beamed. 
We conclude that the magnetospheres of CU Vir, Earth and other planets, maybe also exoplanets, could have similar geometrical and physical characteristics in the regions where the cyclotron maser is generated.
%{\bf We conclude that the tangent plane geometry proposed for the AKR and subsequent refractionof the cyclotron maser radiation in a denser magnetospheric plasma can be applied to theCU Vir case.}
In addition, the pulses are perfect "markers" of the rotation period. This has given us for the first time the possibility to measure with extraordinary  accuracy the spin down of a star on or near the main sequence.
\end{abstract}

\keywords{stars: chemically peculiar --- stars: individual (CU Virginis)
--- stars: magnetic field  
--- polarization --- radiation mechanisms: non-thermal  --- masers}

\section{Introduction}

Among Magnetic Chemically Peculiar (MCP) stars (see \citet{wol83} for an extensive description of this class),  \cu\, (HD124224) is one of the most intriguing. It is a rapidly rotating star, with one of the shortest rotational periods of this class ($P_\mathrm{rot}\approx 0.52^\mathrm{d}$). According to the Oblique Rotator Model \citep{bab49}, it is characterized by a magnetic field with a mainly dipolar topology, with a  strength of $B_\mathrm{p}=3000\pm200$ gauss at the pole, whose axis is tilted of an angle $\beta=74\degr\pm3\degr$ with respect to the rotational one, which in turn subtends an angle $i=43\degr\pm7\degr$ \citep{tri00} to the direction of the Earth. 
The magnetic field is also responsible for the inhomogeneous distribution of the chemical elements in the stellar photosphere, giving rise to photometric and spectroscopic variability. 
%{\bf For \cu, the magnetic field strength at the pole is $B_\mathrm{p}=3000\pm200$ gauss, the inclination of the rotational axis to the direction of the Earth is $i=43\degr\pm7\degr$ and the magnetic axis obliquity $\beta=74\degr\pm3\degr$ \citep{tri00}.}

\cu\, shows continuum radio emission \citep{leo94, leo04}, explained as being due to gyrosynchrotron process from electrons spiraling in a corotating magnetosphere. According the 3D model developed by \citet{tri04} for MCP stars, a radiatively-driven wind interacts with the magnetosphere, dividing it into three regions, namely the "inner", "middle" and "outer" magnetosphere. In the "inner" equatorial zone, the kinetic energy density of the wind particle never exceeds the magnetic one, $B^2/8\pi>(1/2)\rho v^2$,  generating a dense equatorial belt. In the outer region, the path of the ionized wind from the polar cups is first traced by the dipolar field, then,   where $B^2/8\pi<(1/2)\rho v^2$, it is mainly radial. In the thin middle magnetosphere, particles evaporating from a ring around the magnetic poles propagates along the dipolar field lines, breaking them at the Alfv\'en radius ($B^2/8\pi\approx(1/2)\rho v^2$), generating current sheets where particle acceleration occurs. The energetic particles spread into the middle magnetosphere, emitting by the gyrosynchrotron mechanism. For \cu, a weak wind with a mass-loss rate of the order of $10^{-12}\,\mathrm{M}_\odot\,\mathrm{year}^{-1}$ can explain the almost flat spectra and rotational modulation. 

The distinctive behaviour of \cu\, is the presence of pulses of coherent, 100\% circularly polarized, highly directive radio emission at 1.4 GHz \citep{tri00}. Pulses are observed two times every stellar rotation, when the magnetic axis of the dipole is perpendicular to the line of sight. This process has been observed for over ten years \citep{tri00, tri08, rav10}, and it is interpreted as Electron Cyclotron Maser Emission (ECME) originating above the magnetic pole \citep{tri00, tri08}. In the framework of the previous model, the relativistic electrons in a converging flux tube reflect outwards when they arrive close to the star, some fraction of them falling into the stellar photosphere. This causes a loss cone anisotropy which is responsible for the maser emission. 

Another important aspect of the two pulses is that, assuming the emission source is fixed, they represent an ideal marker of the stellar rotation. An abrupt variation of the rotational period of \cu\ has been claimed from optical photometric study by \citet{pyp98}. A change of the period from the radio pulses study has been measured with great accuracy by \citet{tri08}.

In order to study the spectral and temporal behaviour of the pulses, the magnetospheric plasma, the spin down of this star, we observed \cu\, with the EVLA with unprecedented spectral and temporal sensitivity in two bands around the wavelength of 18 cm.

\begin{figure*}
\includegraphics[width=16cm]{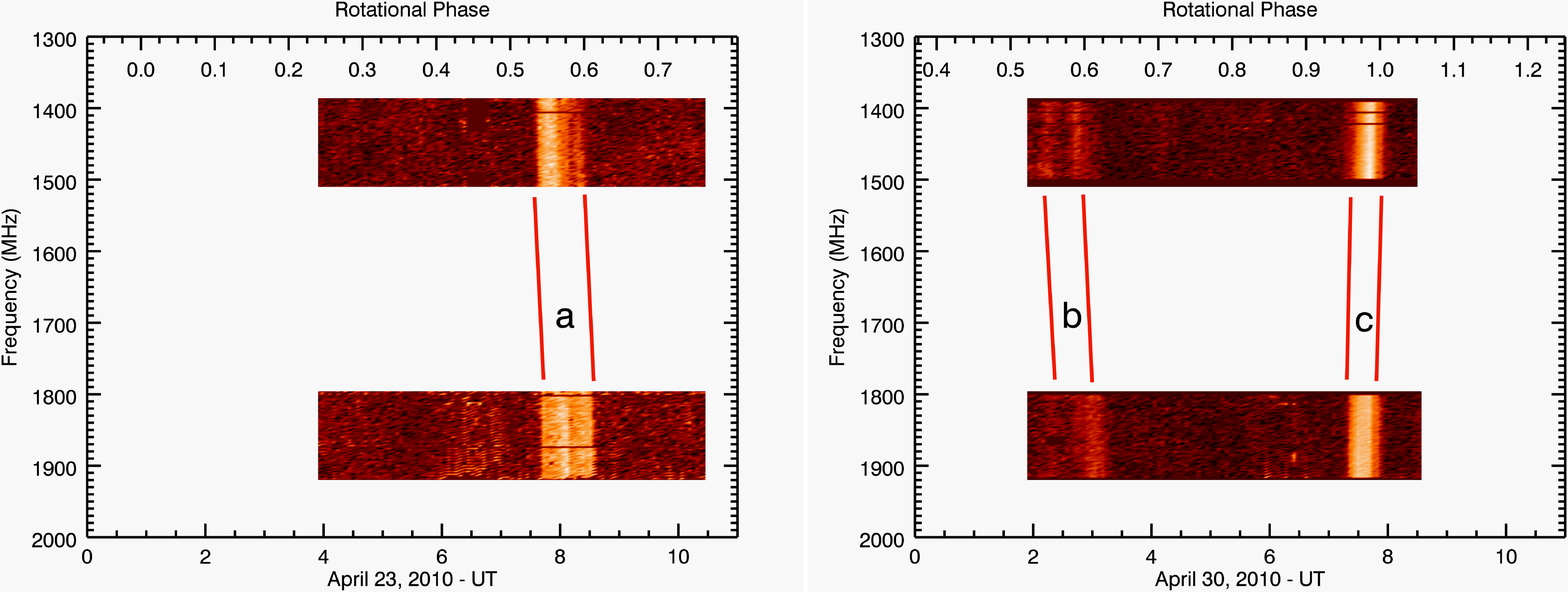}
\caption{Dynamical spectra of the circular polarization (Stokes V) flux density of  CU Vir; the corresponding heliocentric rotational phase is also shown. Left panel) first day: a broad peak, labeled "a", centered at 8 UT is visible; Right panel) second day: two peaks, one at 3 UT, "b", one at 7:30 UT, "c", are visible. Peak "a" and "b" appear at the same rotational phase (0.5-0.6); The peak "c" appears at phase 0.95-1. Vertical lines show that the peaks at $\phi\approx 0.55$ occur early at low frequency; conversely, the peak at $\phi\approx 0.95$ occurs early at high frequency.}
\label{spe}
\end{figure*}

\section{Observations and Data Reduction}
The observations were carried out with the EVLA \citep{per11} in D configuration, in two days, on April 23rd, from 03:36 to 10:35 UT, and on April 30th 2010, form 01:39 to 08:38 UT. The date of the observations have been coordinated in order to cover the whole rotational period of \cu\, (about 12 hours) with two 7-hours slots, with a minimum overlap of rotational phase. We used all the 22 telescopes equipped with the L-band receivers, tuned in two bands centered at 1452 and 1860~MHz respectively, each one with a bandwidth of 128~MHz split into 64 channels, with a frequency resolution of 2~MHz, in full polarization mode. The acquisition on \cu\, was alternated with those on the point-like source \object{J1354-0206}, used as phase calibrator, with a duty cycle of 
$1^{\mathrm m}30^{\mathrm s}-9^{\mathrm m}30^{\mathrm s}-1^{\mathrm m}30^{\mathrm s}$. 
The amplitude scale was determined by observations of the primary calibrator \object{3C286} (F$_{1.4}=14.58$, F$_{1.8}=12.9$~Jy). Data reduction and editing were performed by using the 
\textsc{Common Astronomy Software Applications} package (\textsc{casa}). 
Data inspection revealed much radio interference in both bands, occurring in certain blocks of time and of channels. Bad data have been edited first in the scans of the two calibrators. Bandpass correction has been eventually determined by using the visibilities of \object{3C286}, then flux densities of \object{J1354-0206} have been obtained (F$_{1.4}=774$, F$_{1.8}=854$~mJy). Complex gains have then been applied to the data of \cu\, which have been edited after calibration. Stokes I and V maps of the regions around the phase calibrator and the target have been obtained for the two days. 

In order to get dynamical spectra of our target, visibilities of \cu\, have been imported into the \textsc{Astronomical Image Processing System} (\textsc{aips}), where the task \textsc{dftpl} computes the direct Fourier Transform at a given position in the sky as a function of the time. Stokes V flux densities have been computed for all frequencies, with time and spectral resolution of 2~MHz and 4\,min respectively, giving an rms of about 1~mJy, as expected. 

No Stokes I or V variability has been found for \object{J1354-0206}, neither for the several sources close to \cu.

\begin{figure}[h]
\includegraphics[width=8cm]{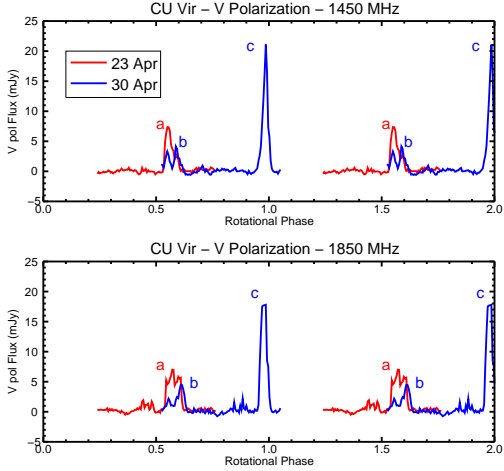}
\caption{Frequency average of the polarized emission as a function of the rotational phase for the two bands.}
\label{phase}
\end{figure}

\begin{figure}
\includegraphics[width=8cm]{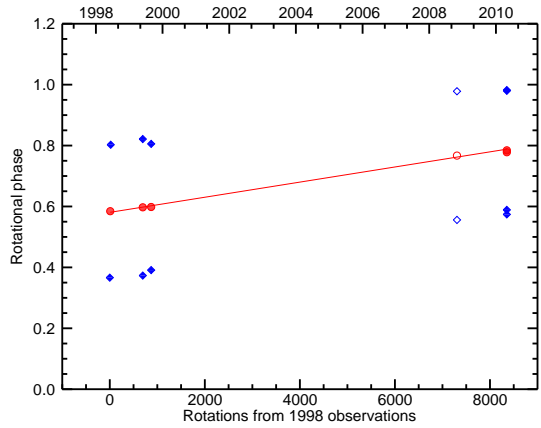}
\caption{
Phases of the peaks at the different epochs of observations, as a function of the number of rotations of the star from the first observation. Years are also shown at the top. The blue diamonds refer to the main emission beams, red dots to the midpoints. From the slope of the linear fit the increment of the rotational period can be inferred.}
\label{delay}
\end{figure}

\begin{figure}
\includegraphics[width=8cm]{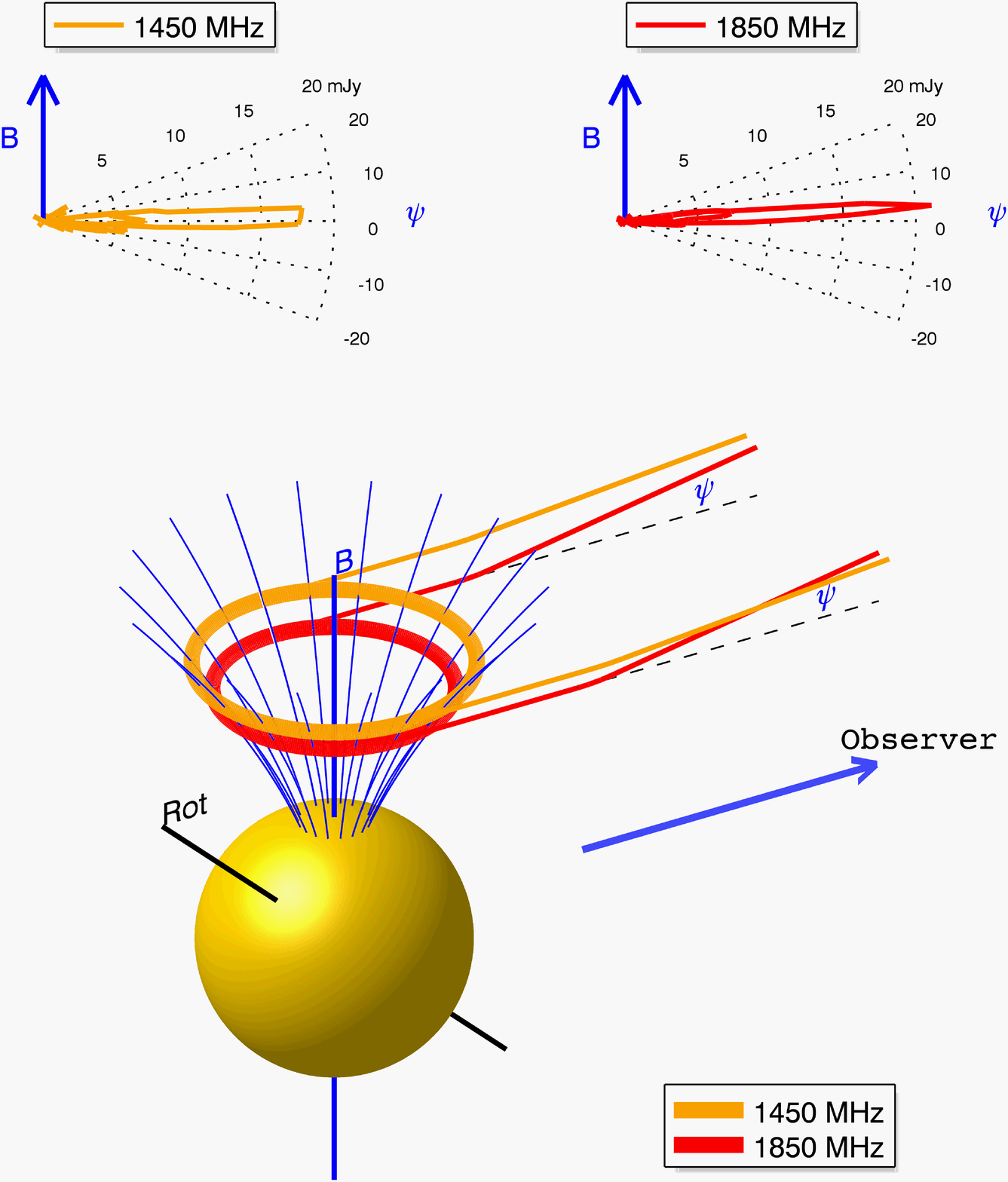}
\caption{Top panel: polar diagrams of the pulses, showing their directivity. 
At 1450~MHz the emission is almost perpendicularly to the magnetic axis $B$; at 1850~MHz there is a deviation of about $4\degr$. 
Bottom panel: picture of the proposed model: maser amplification occurs in annular rings around the pole, at height determined by the gyrofrequency of the emitting electrons; the emission occurs tangentially, then refraction in the higher density region (the "cold torus" in the inner magnetosphere) deviates the radiation upward; this effect is bigger at higher frequency. During the rotation, the pulses are observed from the Earth (supposed to be toward the right) two times, at slightly different moments at different frequencies.}
\label{sphere}
\end{figure}

\section{The Cyclotron Maser}
Dynamical spectra are shown in Fig.~\ref{spe}. Heliocentric rotational phases, computed by using the ephemeris by \citet{pyp98}, indicate that the peak "a" of April 23rd and the "b" of April 30th occur at the same phase ($\phi\approx 0.55$), when the star is oriented in space in the same way relatively to the Earth. Fig.~\ref{phase} shows the polarized flux average in the two bands as a function of the rotational phase, stressing the coincidence in phase of "a" and "b" events and a significant level of variability between the two observations. All the three events last approximately 1 hour each; "c" exhibits a single peak of flux levels of about 20 mJy; "a" and "b" lower and broader emission, with two components in the "b" event. The phase difference between "a" (or "b") and "c" events is about 0.4. This allows to recognize the same peaks of emission reported by \citet{tri00}, \citet{tri08} and \citet{rav10}. 
The difference with the previous observations is the rotational phase of each peak, not the difference of phase between them. The variability of the pulse shape, as well as the amplitude, could be due to some instability of the emission region, as proposed by \citet{rav10}. Many parameters enter in the growth rate $\omega$ of the cyclotron maser \citep{mut07} such as electron density, velocity distribution ($f$) and its anisotropy $(\partial f/\partial v_\bot)$, plasma frequency ($\nu_{\mathrm p}$) and others. Since the growth rate is a non linear function of them, and the intensity of an unsaturated maser is $I\propto \exp(\omega)$, even a small variation of one of the above parameters can result in a significant variation of the emerging flux.

\subsection{The change of the rotational period}
An abrupt period change of 2.18 seconds, occurring probably in 1985, has been claimed by \cite{pyp98} from the analysis of photometric data; a new spin down of about 1 second has been inferred by  \citep{tri08} from the phase shift of the coherent radio emission between 1998 and 1999. Now, with a new set of points after 12 years, and with the inclusion of the \citet{rav10} results, the spin down is confirmed and the increment of the rotational period $\Delta P$ is determined with great accuracy by fitting the midpoints of the two peaks (circles and line in  Fig.~\ref{delay}):  
$\Delta P=1.12 \pm 0.03$~s. 
If we plot the phase of the peaks as a function of the epoch (Fig.~\ref{delay}), we see that they appear with a regularly increasing delay, indicating that the rotational period of \cu\, has changed. Assuming that the change of period occurred close to the observations of 1998, we determine new ephemeris of \cu: 
$$
HJD = 245 0966.3601 + 0.52071601 E
$$
for $JD>245 0966$, nominally June 1998. It is worthwhile to stress that this method of period determination gives a relative error of $7\times 10^{-7}$.

\subsection{The direction of the beams}
After computing the actual rotational phases $\phi$ with the new ephemeris, it is possible to get the orientation of the star and its magnetosphere, then the direction of emission with respect to the dipole axis. Assuming an inclination $i=43\degr$ of the rotational axis and an obliquity $\beta=74\degr$ of the magnetic axis \citep{tri00}, the angle $\psi$ between the line of sight and the axis of the dipole is given by:
\[
\sin \psi=\sin \beta \sin i \cos(\phi-\phi_0)+\cos \beta \cos i;
\]
here $\phi_0=0.08$ is the phase delay of the magnetic curve with respect to the light curve \citep{bor80}. The maximum visibility of the North magnetic hemisphere is at $\phi=0.58$, while at $\phi=0.38$ and $0.78$ the magnetic axis lies in the plane of the sky. 

By looking at the radio emission as a function of the inclination of the magnetic axis, we can see how it escapes from the emitting region, i.e. we can study its directivity. The direction of the emission beams at 1450 and 1850 MHz relatively to the dipole axis are shown in Fig.~\ref{sphere}, top panel. Beams at higher frequency are emitted at $\psi\approx 4\degr$, while at lower frequency at $\psi\approx 0\degr$, as the visibility of the peaks in Fig.~\ref{spe} indicates. The emitting beams subtend an angle $\Delta \psi \la10\degr$ HPBW.

The cyclotron maser hypothesis in \cu\ has already been proposed by \citet{tri00,tri08}. To be possible, the emission must originate in a region of relatively strong magnetic field ($\nu_{\rm{B}}>>\nu_{\rm{P}}$), and the plasma density in region above the magnetic pole must be relatively low. Following the ECME theory \citep{mel82}, the direction of emission forms an angle $\theta \la 90\degr$ ($\cos\theta\approx v/c$, where $v$ is the upward velocity of the emitting electrons) to the magnetic field direction with an aperture $\Delta\theta\approx v/c$), forming an hollow cone. The frequency of the radiation is $\nu \ga s\;\nu_{\rm{B}}$, where $s$ is the harmonic number. It is probably the second harmonic which is the most efficient one, since emission at the first harmonic is subject to gyromagnetic absorption of nearby thermal plasma at $s=2$, and for a higher harmonic number the maser has less efficient amplification. Assuming $s=2$, the emission occurs where $\nu=2\times 2.8 B/{\rm{G}}$~MHz. For our two observing frequencies of 1450 and 1850 MHz, the corresponding magnetic field is respectively 259 and 330~G and, with a polar field strength $B_{\rm{pol}}=3000$~G, the observed emission arises respectively at $R=2.26$ and $2.09\;R_{\ast}$ from the center of the star.

There are however some problems with the hollow cone beaming model. In fact, if the emission arises from a ring around the polar axis and if the emission is in direction $\theta\la90\degr$, we should observe much broader peaks, since the inclination of the magnetic field lines with respect to the line of sight  changes in the emitting rings (see Fig.~4 of \citet{tri08}). So the open questions are: 

-why do we not observe broad peaks?

-why does the angle of emission change with the frequency?

\section{"Auroral" emission}
The cyclotron maser of \cu\ shares with the radio emission from planets some characteristics, suggesting a common scenario. There are five planets with magnetospheres in the Solar System (Earth, Jupiter, Saturn, Uranus and Neptune) known to show intense radio emission at low frequency (from 10 kHz up to few MHz), originating above the high magnetic latitude auroral regions \citep{zar98}. This emission is non-thermal, with high brightness temperature exceeding $10^{15}$K, 100\% circularly polarized, highly beamed. All those characteristics indicate that the emission process is cyclotron maser. Except for the case of the Jovian decametric emission (DAM) at $10-40$ MHz, which is due to the Io-magnetosphere interaction, the auroral emissions are controlled by the solar wind flowing into the open field lines above the polar caps. The analogy with \cu\ is unexpected, including the "pulsar like" behaviour due to the rotation of an oblique dipole axis. 

Recent observations of the terrestrial Auroral Kilometric Radiation (AKR), performed with the four-spacecraft Cluster array, have allowed the location of the sources from which the emission originates \citep{mut03} and the determination of its the angular beaming characteristics \citep{mut08}. The emission seems to be the sum of several elementary bursts of radiation generated at a height where the gyrofrequency of the electrons corresponds to the magnetic field (at $0.5-2 R_{\oplus}$), in a ring around the magnetic axis where the density is very low (the so-called "auroral cavity"). Unlike the Jovian DAM, where the emission is beamed in a thin ($\Delta\theta\approx  1\degr$) hollow cone of half-angle $\theta\la 90\degr$ \citep{dul70}, the AKR is radiated within $\pm15\degr$ from the "plane tangent to the source's magnetic latitude circle and containing the local magnetic field vector" \citep{mut08,lou96}. The important point we want to stress is that the emission is emitted tangentially to the auroral ring. 

During the propagation in the close ambient medium, the AKR is deviated upwards because of the refractive index of the dense magnetospheric plasma.\\

\subsection{A new model for CU Vir}
The results of AKR suggest that the ECME of \cu\ could come from the same geometry. We propose a new scenario, where the masing emission does not follow the "hollow cone model", instead the "tangent plane beaming model". Fig. \ref{sphere} shows a sketch of the Cyclotron Maser emission from \cu. Electrons traveling back from the acceleration region located at Alfv\'en radius $R_{\rm{Alf}}=15\;R_{\ast}$ \citep{let06} emit at locations where the magnetic field strength corresponds to the first or second gyromagnetic harmonic ($s=1,2$). These define annular rings above the magnetic pole. The amplification occurs tangentially, so that it is seen from Earth to came from the two extreme points of the rings. They would be seen simultaneously at any frequency as the star rotates, in the absence of subsequent refraction. 

Our 3-D model \citep{tri04,let06} foresees the existence of a cold torus around the magnetic equator. The absorption of this optically thick material can account for the rotational modulation of the continuous gyrosynchrotron emission, in particularly at high frequency radiation ($\nu>15$~GHz), emitted exactly in the same region where the cyclotron maser originates. From the results of the simulation, the number density of the cold torus is $N_{\rm{torus}}\approx10^9\rm{cm}^{-3}$, with a temperature $T_{\rm{torus}}\approx10^4\rm{K}$. 

We now consider the case of $s=2$, and will give some consideration for the case $s=1$. During the propagation, the maser radiation crosses the torus. The plasma of the torus, permeated by a magnetic field $B\approx260$ and $330$~G (the levels of the propagation of the two frequency bands) has a refractive index in the extraordinary mode given by: 
\[
n_{\rm{refr}} \approx \sqrt{1 - \frac{\nu_{\rm{P}}^2}
{\nu(\nu-\nu_{\rm{B}})}  }
\]
In the following, we will give an order of magnitude estimate of the deviation of the emission (angle $\psi$) with respect to the original, horizontal, ray path. The refractive index inside the torus, assuming an average $N_{\rm{torus}}$ as above, but with higher density in the inner part, and the local field strength, can easily reach values of $n_{\rm{refr}} \approx 0.98$ and $0.95$ for $\nu=1450$ and $1850$~MHz respectively. This means that the maximum deviation toward high values of the angle $\psi$ occurs at $1850$~MHz, in agreement with our observations. The actual angle $\psi$ can be obtained in a simply way from the Snell law if the angle of incidence with the interface between the two media at different refractive index, i.e. the surface defined by the field lines confining the cold torus, is known. In the case that this angle is $60\degr$, $\psi\approx 10\degr$ at 1850~MHz and $\psi\approx 5\degr$ at 1450~MHz, accounting for the observed difference (see Fig.~\ref{sphere}). 

In the case $s=1$, this simple computation foresees bigger deviations upwards, not in agreement with the observations. However, a more detailed numerical computation is needed in order to better constrain the actual harmonic number $s$ and get more precise indications on the physical conditions of the emitting plasma and of the surrounding region. 

\section{Conclusion and future perspective}
The observations reported here  of \cu\ confirm the presence of a steady cyclotron maser operating over more than 10 years in the magnetosphere of this star. Our model proposes that the process is triggered by the continuous injection of high energy electrons flowing from the Alfv\'en region, about 15 stellar radii from the star, to the inner parts of the magnetosphere. Those electrons are also responsible for the gyrosynchrotron emission. The overall mechanism is powered by the radiation driven wind interacting with the magnetic field. 

The high temporal and spectral resolution of the current EVLA allowed us to get dynamical spectra over two broad spectral regions.  The observations have been carried in two days, covering almost the whole rotational period of \cu. This allowed us to measure with great accuracy the delay of the peaks of emission, and so to determine with unprecedented precision ($\approx10^{-7}$) the rotational period of the star, confirming the spin down of this young star. Further observations are needed in order to monitor the rotation of \cu , since the emission peaks are believed to be "perfect markers" of this stellar clock. 

There is an impressive similarity between the cyclotron maser of \cu\ and the coherent radio emission from planets. The recent observations of the Auroral Kilometric Radiation from the terrestrial magnetosphere and the new interpretation on the basis of the "tangent plane beaming model" gave the opportunity to formulate a new model for \cu, where the emission is "auroral-type", emitted tangentially to the auroral circles. 

The analysis of the dynamical spectra permitted us to clearly see that the pulses occurred at slightly different times inside the frequency band. In the framework of the new model, this could be due to refractive effects due to the propagation of the radiation inside a denser magnetized plasma surrounding the star in the magnetic equatorial belt. This is a further indication of the existence of a "cold torus", which also acts as an absorber of the gyrosynchrotron continuous emission at centimeter wavelengths. 

In the near future the EVLA will increase its maximum bandwidth up to 2~GHz. This will allow us to extend the dynamical spectra in a broader spectral region and to better define the low and high limits of the cyclotron maser. The time visibility of the peaks as a function of the frequency will allow us to study and model with great precision the propagation of the radiation in the magnetospheric plasma surrounding the star. \cu\ is an excellent astrophysical laboratory for plasma physics in stellar and planetary environment. 

\acknowledgments
We thank the NRAO scheduler who coordinated the times of the two observations to cover almost the whole rotational period of \cu. The EVLA is a facility of the National Radio Astronomy Observatory which is operated by Associated Universities, Inc. under cooperative agreement with the National Science Foundation.

{\it Facilities:} \facility{EVLA}.


\begin{thebibliography}{}
\bibitem[\protect\citeauthoryear{Babcock}{1949}]{bab49}
	Babcock, H.W. 1949, Observatory, 69, 191
\bibitem[\protect\citeauthoryear{Borra \& Landstreet}{1980}]{bor80}
	Borra, E.F., \& Landstreet, J.D. 1980, \apjs, 42, 421
\bibitem[\protect\citeauthoryear{Dulk}{1970}]{dul70}
	Dulk, G.A. 1970, \apj, 159, 671
\bibitem[\protect\citeauthoryear{Leone et al.}{1994}]{leo94}
	Leone, F., Trigilio, C., \& Umana, G. 1994, \aap, 263, 908
\bibitem[\protect\citeauthoryear{Leone et al.}{2004}]{leo04}
	Leone, F., Trigilio, C., Neri, R., \& Umana, G. 2004, \aap, 423, 1095
\bibitem[\protect\citeauthoryear{Leto et al.}{2006}]{let06}
	Leto, P., Trigilio, C., Buemi, C. S., Umana, G., \& Leone, F. 2006, \aap, 458,831
\bibitem[\protect\citeauthoryear{Louarn \& Le Qu\'eau}{1996}]{lou96}
	Louarn, P. \& Le Qu\'eau, D. 1996, \planss, 44, 211
\bibitem[\protect\citeauthoryear{Melrose \& Dulk}{1982}]{mel82}
	Melrose, D.B., \& Dulk, G.A. 1982, \apj, 207, 341
\bibitem[\protect\citeauthoryear{Mutel et al.}{2003}]{mut03}
	Mutel, R.L., Gurnett, D. A., Christopher, I. W., Pickett, J. S., \& Schlax, M. 2003, \jgr, 108, 1398
\bibitem[\protect\citeauthoryear{Mutel et al.}{2007}]{mut07}
	Mutel, R.L., Peterson, W. M., Jaeger, T. R., \& Scudder, J. D. 2007, \jgr, 112, 7211
\bibitem[\protect\citeauthoryear{Mutel et al.}{2008}]{mut08}
	Mutel, R.L., Christopher, I.W., \& Pickett, J.S. 2008, \grl, 35, L07104.
\bibitem[\protect\citeauthoryear{Perley et al.}{2011}]{per11}
	Perley, R.A., Chandler, C.J., Butler, B.J., Wrobel, J.M. 2011, 
	\apjl, in press 
\bibitem[\protect\citeauthoryear{Pyper et al.}{1998}]{pyp98}
	Pyper, D.M., Ryabchikova, T., Malanushenko, V., Kuschnig, R., Plachinda, S., \& Savanov, I. 1998, 
	\aap, 339, 822
\bibitem[\protect\citeauthoryear{Ravi et al.}{2010}]{rav10}
	Ravi, V., et al. 2010, \mnras, 408, L99
%\bibitem[\protect\citeauthoryear{Stibbs}{1950}]{sti50}
%	Stibbs, D. W. N. 1950,  \mnras, 110, 395
\bibitem[\protect\citeauthoryear{Trigilio et al.}{2000}]{tri00}
	Trigilio, C., Leto, P., Leone, F., Umana, G., \& Buemi, C. 2000, \aap, 362, 281
\bibitem[\protect\citeauthoryear{Trigilio et al.}{2004}]{tri04}
 	Trigilio, C., Leto, P., Umana, G., Leone, F., \& Buemi, C.S. 2004, \aap, 418, 593
\bibitem[\protect\citeauthoryear{Trigilio et al.}{2008}]{tri08}
	Trigilio, C., Leto, P., Umana, G., Buemi, C.S., \& Leone, F. 2008, \mnras, 384, 1437
\bibitem[\protect\citeauthoryear{Wolff}{1983}]{wol83}
	Wolff, S.C. 1983, NASA SP-46
\bibitem[\protect\citeauthoryear{Zarka}{1998}]{zar98}
	Zarka, P. 1998, \grl, 103, 20159
\end{thebibliography}
\end{document}